\documentclass[superscriptaddress,preprintnumbers,amssymb,twocolumn]{revtex4}
\usepackage{graphicx}
\usepackage[latin1]{inputenc}
\usepackage{epsfig}
\usepackage{amsmath}
\usepackage{amsfonts}
\usepackage{amssymb}

\def\lsim{\:\raisebox{-0.5ex}{$\stackrel{\textstyle<}{\sim}$}\:}
\def\gsim{\:\raisebox{-0.5ex}{$\stackrel{\textstyle>}{\sim}$}\:}

\begin{document}

\title {The 125 GeV Higgs in the context of four generations with 2 Higgs doublets}
\author{Michael Geller}
\email{mic.geller@gmail.com}
\affiliation{Physics Department, Technion-Institute of Technology, Haifa 32000, Israel}
\author{Shaouly Bar-Shalom}
\email{shaouly@physics.technion.ac.il}
\affiliation{Physics Department, Technion-Institute of Technology, Haifa 32000, Israel}
\author{Gad Eilam}
\email{eilam@physics.technion.ac.il}
\affiliation{Physics Department, Technion-Institute of Technology, Haifa 32000, Israel}
\affiliation{Center for High Energy Physics, Indian Institute of Science, Bangalore 500 012,
India}
\author{Amarjit Soni}
\email{soni@bnl.gov}
\affiliation{Theory Group, Brookhaven National Laboratory, Upton, NY 11973, USA}

\date{\today}

\begin{abstract}
We interpret the recent discovery of a 125 GeV Higgs-like state in the context
of a two Higgs doublets model with a heavy 4th sequential generation of fermions,
in which one Higgs doublet couples only to the 4th generation fermions, while the
second doublet couples to the
lighter fermions of the 1st-3rd families. This model
is designed
to accommodate the apparent heaviness of the 4th generation fermions and to
effectively address the low-energy phenomenology
of a dynamical electroweak symmetry breaking scenario.
The physical Higgs states of the model
are, therefore, viewed as composites
primarily of the 4th generation fermions.
We find that the lightest Higgs, $h$, is a good candidate for
the recently discovered 125 GeV spin-zero particle,
when $\tan\beta \sim {\cal O}(1)$, for typical 4th generation fermion masses
of $M_{4G} = 400 -600$ GeV,
and with a large $t - t^\prime$ mixing in the right-handed
quarks sector. This, in turn, leads to
${\rm BR}(t^\prime \to t h) \sim {\cal O}(1)$,
which drastically changes the $t^\prime$ decay pattern.
We also find that, based on the current Higgs data, this two Higgs doublet
model generically predicts an enhanced production rate (compared to the SM) in the
$pp \to h \to \tau \tau$ channel and a reduced $VV \to h \to \gamma \gamma$ and
$p \bar p/pp \to V \to  hV \to V bb$ ones.
Finally, the heavier CP-even Higgs
is excluded by the current data up to $m_H \sim 500$ GeV, while
the pseudoscalar state, $A$, can be as light as 130 GeV.
These heavier Higgs states and the expected deviations
from the SM in some of the Higgs production channels
can be further excluded or discovered
with more data.
\end{abstract}


\maketitle

\section{Introduction}

The LHC has recently observed a new scalar particle with a mass around $\sim 125$ GeV that could be consistent with the Higgs boson of the Standard Model (SM) \cite{CMSHiggs,ATLASHiggs}.
In addition, a study of the combined Tevatron data has revealed a smaller broad excess corresponding to a mass between 115 GeV and 135 GeV \cite{TevatronHiggs} which is consistent with this
LHC discovery. With more data collected, the LHC is expected to be able to
unveil the detailed properties of the new scalar particle and verify its nature.

From the theoretical side, there has been a collective effort in the past decades in the search
for new physics beyond the SM, that can address some of the fundamental unresolved questions
in particle physics. One simple candidate that was extensively studied in the past several years
is the so called SM4 (also referred to as
``naive" or ``simple" SM4); the SM with a fourth sequential generation of fermions (for reviews see \cite{sher0,hou2009-rev,SM4proc,khlopov}).
This simple extension of the SM was studied for addressing
some of the challenges in particle physics, such as:
the hierarchy problem \cite{DEWSB,holdom-new,hung1,ghnew},
the origin of matter - anti matter asymmetry in the universe \cite{baryo-ref,fok} and the
apparent anomalies in flavor physics \cite{SAGMN08,SAGMN10,ajb10B,buras_charm,4Gflavor}.

Unfortunately, the recent LHC searches for the SM4 heavy 4th generation quarks
have now pushed the exclusion limits to $\sim 550$ GeV for the $t'$ and $\sim 600$ GeV for the $b'$ \cite{LHC_4G}, which is on the border of their perturbative regime.
Moreover, the SM4 Higgs was already excluded in the mass range $120-600$ GeV by the 2011 data \cite{SM4-Higgs-bound} when $m_{\nu_4}>m_h/2$. Thus, the above reported discovery
 of a light Higgs with a mass around 125 GeV is not compatible with the SM4
 that includes a heavy 4th generation neutrino with a mass $m_{\nu_4} \gsim 100$ GeV, i.e.,
 with a mass larger than the current lower bound on $m_{\nu_4}$ \cite{PDG}.
 In fact, it was further pointed out recently in \cite{Lenz,Lenz2,Nir,Lenz3}
 that the interpretation of the measured Higgs signals
 is not consistent with the SM4 also for the case $m_{\nu_4} < m_h/2$.
In particular, in the SM4, the leading gluon-fusion light Higgs production mechanism
is enhanced by a factor of $\sim 10$
due to the contribution of diagrams with $t^\prime$ and $b^\prime$ in the loops \cite{Kribs_EWPT}, which
in general leads to larger signals (than what was observed at the LHC) in the $h \to ZZ/WW/\tau\tau$ channels. However, if the 4th generation masses are of ${\cal O}(600)$ GeV,
then the decay channels $h\to ZZ/WW$ are suppressed due to NLO corrections \cite{Passarino,Passarino2}, and the exclusion of the SM4 is based mainly on the $\tau\tau$ channel.
In the $h\to \gamma\gamma$ channel there is also a substantial suppression of ${\cal O}(0.1)$
due to (accidental) destructive interference in the loops \cite{Kribs_EWPT,He3}
and another ${\cal O}(0.1)$ factor due to NLO corrections \cite{Passarino,Passarino2}.
When $\nu_4$ is taken to be light enough so that $Br(h\to \nu_4 \nu_4)$ becomes ${\cal O}\left(1\right)$,
then the $\gamma\gamma $ channel becomes further suppressed to the level that the observed excess can no longer be accounted for \cite{Lenz2}. For a recent comprehensive analysis of the SM4
status in light of the latest Higgs results and electroweak precision data (EWPD), we refer
the reader to \cite{12091101}.

However, as was noted already 20 years ago \cite{Luty} and more recently
in \cite{hashimoto1,4G2HDM}, if heavy 4th generation fermions are viewed as the agents
of electroweak symmetry breaking (and are, therefore, linked to
strong dynamics at the nearby TeV-scale), then more Higgs particles are expected at
the sub-TeV regime. In this case, the Higgs particles may be
composites of the 4th generation fermions and the low-energy
composite Higgs sector should resemble a two (or more) Higgs doublet framework.
Indeed, the phenomenology of multi-Higgs 4th generation models
was studied recently in
\cite{hashimoto1,4G2HDM,soninew,sher2HDM,sher1,Bern,Gunion,FermCond,gustavo1,gustavo2,hung3,hung2,wise1,valencia,He2,alfonso1}
and within a SUSY framework in \cite{fok,shersusy,dawson,rizzo}, for a review
see \cite{AHEP-rev}. In \cite{Geller} it was also shown
that the current exclusion limits on the SM4 $t^\prime$ and $b^\prime$
could be significantly relaxed if the four generations scenario is embedded
in a 2HDM framework.

Adopting this viewpoint, i.e., that the 4th generation setup should be more adequately described
within a multi-Higgs framework, we will study in this paper the expected Higgs signals of a 2HDM with a 4th generation family, investigating whether the interpretation of the recently measured 125 GeV Higgs properties are consistent with one of the neutral scalars of the 4th generation 2HDM.

\section{2HDM's and 4th generation fermions \label{sec2}}

The 2HDM structure has an inherent freedom of choosing which doublet couples to which fermions. For the three generations 2HDM, three popular setups were suggested which are usually referred to as
type I, type II and type III 2HDM.
In the case where the 2HDM is assumed to underly some form of
TeV-scale strong dynamics mediated by the 4th generation fermions, we expect the Higgs composites
to couple differently to the 4th generation fermions.
This can be realized in a class of 2HDM models named the 4G2HDM, suggested in \cite{4G2HDM}.
Most of our analysis below is performed in this 4G2HDM framework and a comparison to a 2HDM of type II (which also underlies the SUSY Higgs sector)
with and without 4th generations will also be discussed.

Let us recapitulate the salient features of the 2HDM frameworks with a 4th generation
 of fermions (we will focus below on the quarks sector, but a generalization to the leptonic
  sector is straight forward). Assuming a common generic 2HDM potential, the phenomenology of 2HDM's is generically encoded in the texture of the Yukawa interaction Lagrangian.
The simplest variant of a 2HDM with 4th generations of fermions, can
be constructed based on the so called type II 2HDM
(which we denote hereafter by 2HDMII), in which one of the Higgs
doublets couples only to up-type fermions and the
other to down-type ones. This setup ensures the absence of tree-level
flavor changing neutral currents (FCNC)
and is, therefore, widely favored
when confronted with low energy flavor data. The Yukawa terms for the quarks of
the 2HDMII, extended to include the extra
4th generation quark doublet is:
\begin{eqnarray}
\mathcal{L}_{Y}= -\bar{Q}_{L} \Phi_{d} F_d d_{R}
-\bar{Q}_{L} \tilde\Phi_{u} F_u u_{R} + h.c.\mbox{ ,}
\label{eq:LYII}
\end{eqnarray}
where $f_{L(R)}$ ($f=u,d$) are left(right)-handed fermion fields, $Q_{L}$ is the left-handed
$SU(2)$ quark doublet, $F_d,F_u$ are general $4\times4$
Yukawa matrices in flavor space and $\Phi_{d,u}$ are the Higgs doublets:
\begin{eqnarray}
\Phi_i & =\left(\begin{array}{c}
\phi^{+}_i\\
\frac{v_i+\phi^{0}_i}{\sqrt{2}}\end{array}\right),\quad\tilde{\Phi_i}=\left(\begin{array}{c}
\frac{v_i^{*}+\phi^{0*}_i}{\sqrt{2}}\\
-\phi^{-}_i\end{array}\right) ~.
\end{eqnarray}

As mentioned above, motivated by the idea that the low energy scalar degrees of freedom may be the composites of the heavy
4th generation fermions, \cite{4G2HDM} have constructed a new class of 2HDM's, named the 4G2HDM, that can effectively parameterize 4th generation condensation by giving a special status to the 4th family fermions. The possible viable variants of this approach can be parameterized as \cite{4G2HDM}:
\begin{widetext}
\begin{eqnarray}
\mathcal{L}_{Y}= -\bar{Q}_{L}
\left( \Phi_{\ell}F_d \cdot \left( I-{\cal I}_d^{\alpha_d \beta_d}
\right) +
\Phi_{h}F_d \cdot {\cal I}_d^{\alpha_d \beta_d} \right) d_{R}
-\bar{Q}_{L}
\left( \tilde\Phi_{\ell} G_u \cdot \left( I - {\cal I}_u^{\alpha_u \beta_u} \right) +
\Phi_{h} G_u \cdot {\cal I}_u^{\alpha_u \beta_u} \right)
u_{R} + h.c.\mbox{ ,}
\label{eq:LY4G}
\end{eqnarray}
\end{widetext}
where $\Phi_{\ell,h}$ are the two Higgs doublets,
$I$ is the identity matrix and ${\cal I}_q^{\alpha_q \beta_q}$ ($q=d,u$)
are diagonal $4\times4$ matrices defined by
${\cal I}_q^{\alpha_q \beta_q} \equiv {\rm diag}\left(0,0,\alpha_q,\beta_q\right)$.

In particular, in the type I 4G2HDM of \cite{4G2HDM}
(which we will focus on below and which will be denoted hereafter simply as the 4G2HDM),
one sets $\left(\alpha_d,\beta_d,\alpha_u,\beta_u\right)=\left(0,1,0,1\right)$, so that
the ``heavier" Higgs field ($\phi_h$) is assumed to
couple only to the 4th generation quarks,
while the ``lighter" Higgs field ($\phi_\ell$)
is responsible for the mass generation of all other (lighter 1st-3rd generations) fermions.

The Yukawa interactions for these 4G2HDM models in terms of the physical
states were given in \cite{4G2HDM}.
For the lighter CP-even Higgs it reads:
\begin{widetext}
\begin{eqnarray}
{\cal L}(h q_i q_j) &=& \frac{g}{2 m_W} \bar q_i \left\{ m_{q_i} \frac{s_\alpha}{c_\beta} \delta_{ij}
- \left( \frac{c_\alpha}{s_\beta} + \frac{s_\alpha}{c_\beta} \right) \cdot
\left[ m_{q_i} \Sigma_{ij}^q R + m_{q_j} \Sigma_{ji}^{q \star} L \right] \right\} q_j h \label{Sff1}~, \\
\end{eqnarray}
\end{widetext}
where $\alpha$ is mixing angle in the CP-even neutral Higgs sector,
$\tan\beta = v_h/v_\ell$ is the ratio between the VEVs of $\Phi_h$ and $\Phi_\ell$
and $\Sigma^d$, $\Sigma^u$ are
new mixing matrices in the down(up)-quark sectors,
which are obtained after diagonalizing the quarks mass matrices.
These matrices are key parameters of the model, which depend
on the rotation (unitary) matrices of the right-handed down and up-quarks,
$D_R$ and $U_R$, and
on whether $\alpha_q$ and/or $\beta_q$ are ``turned on":
\begin{eqnarray}
\Sigma_{ij}^d &=&  \alpha_d D_{R,3i}^\star D_{R,3j} + \beta_d D_{R,4i}^\star D_{R,4j}~, \nonumber \\
\Sigma_{ij}^u &=&  \alpha_u U_{R,3i}^\star U_{R,3j} + \beta_u U_{R,4i}^\star U_{R,4j} ~.
\label{sigma}
\end{eqnarray}

Thus, as opposed to ``standard" 2HDM's, in the 4G2HDM
some elements of $D_R$ and $U_R$ are physical and
can, in principle, be measured in Higgs-fermion systems.
In particular, inspired by the working assumption of the 4G2HDM
and by the observed flavor pattern in the up and down-quark sectors,
it was shown in \cite{4G2HDM} that, for $\left(\alpha_d,\beta_d,\alpha_u,\beta_u\right)=\left(0,1,0,1\right)$,
the new mixing matrices $\Sigma^d$ and $\Sigma^u$ are
expected to have the following form:
%
\begin{eqnarray}
\Sigma^u &=& \left(\begin{array}{cccc}
0 & 0 & 0 & 0 \\
0 & 0 & 0 & 0 \\
0 & 0 & |\epsilon_t|^2 & \epsilon_t^\star \left( 1- \frac{|\epsilon_t|^2}{2} \right) \\
0 & 0 &  \epsilon_t \left( 1- \frac{|\epsilon_t|^2}{2} \right) & \left( 1- \frac{|\epsilon_t|^2}{2} \right)
\end{array}\right), \label{sigsimple}
\end{eqnarray}
%
and similarly for $\Sigma^d$ by replacing $\epsilon_t \to \epsilon_b$.
The new parameters $\epsilon_t$ and $\epsilon_b$ are free parameters of the model
that effectively control the mixing between the 4th generation and the
3rd generation quarks. We therefore expect $\epsilon_b << \epsilon_t$, so that
a natural choice for these parameter would
be $\epsilon_b \sim {\cal O}(m_b/m_{b^\prime})$
and $\epsilon_t \sim {\cal O}(m_t/m_{t^\prime})$ (see also \cite{4G2HDM}).
In what follows we will thus set
$\epsilon_b = 0$ and vary the $t - t^\prime$ mixing parameter
in the range $0 < \epsilon_t < 0.5$.

\section{2HDM's and the 125 GeV Higgs Signals}
\begin{figure}[htb]
\begin{center}$
\begin{array}{cc}
\epsfig{file=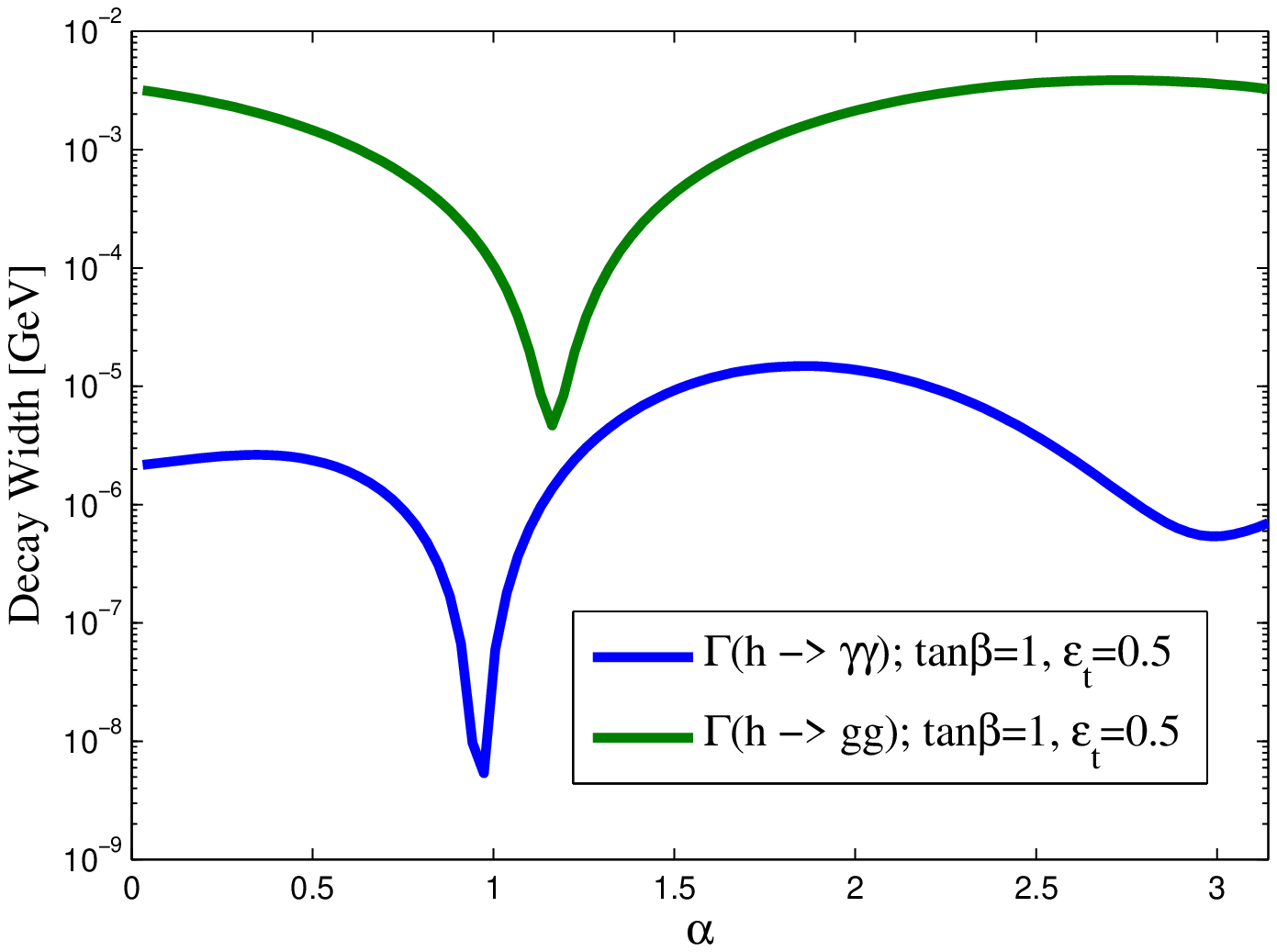,height=6.0cm,width=6.5cm,angle=0}\\
\epsfig{file=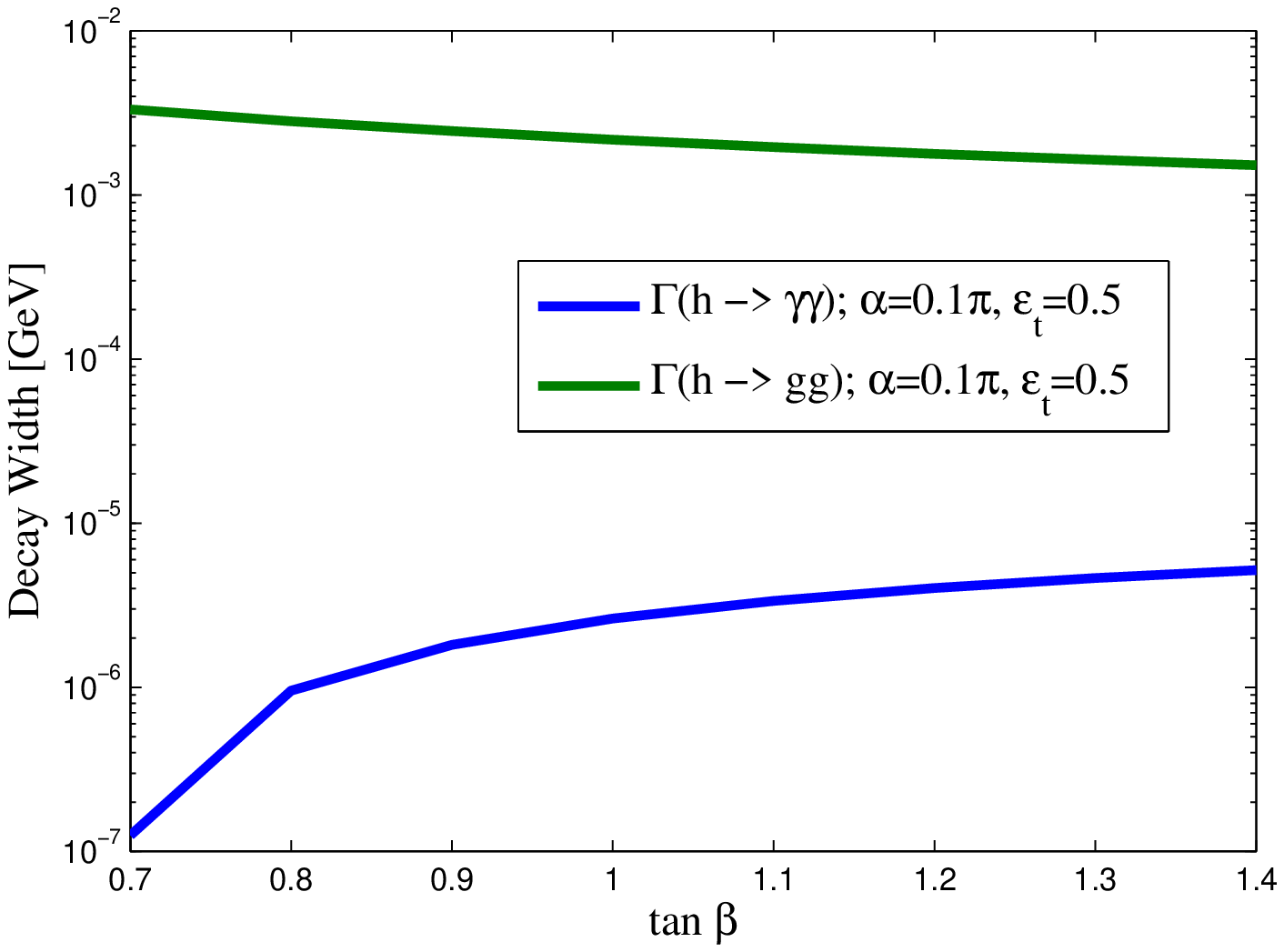,height=6.0cm,width=6.5cm,angle=0}\\
\epsfig{file=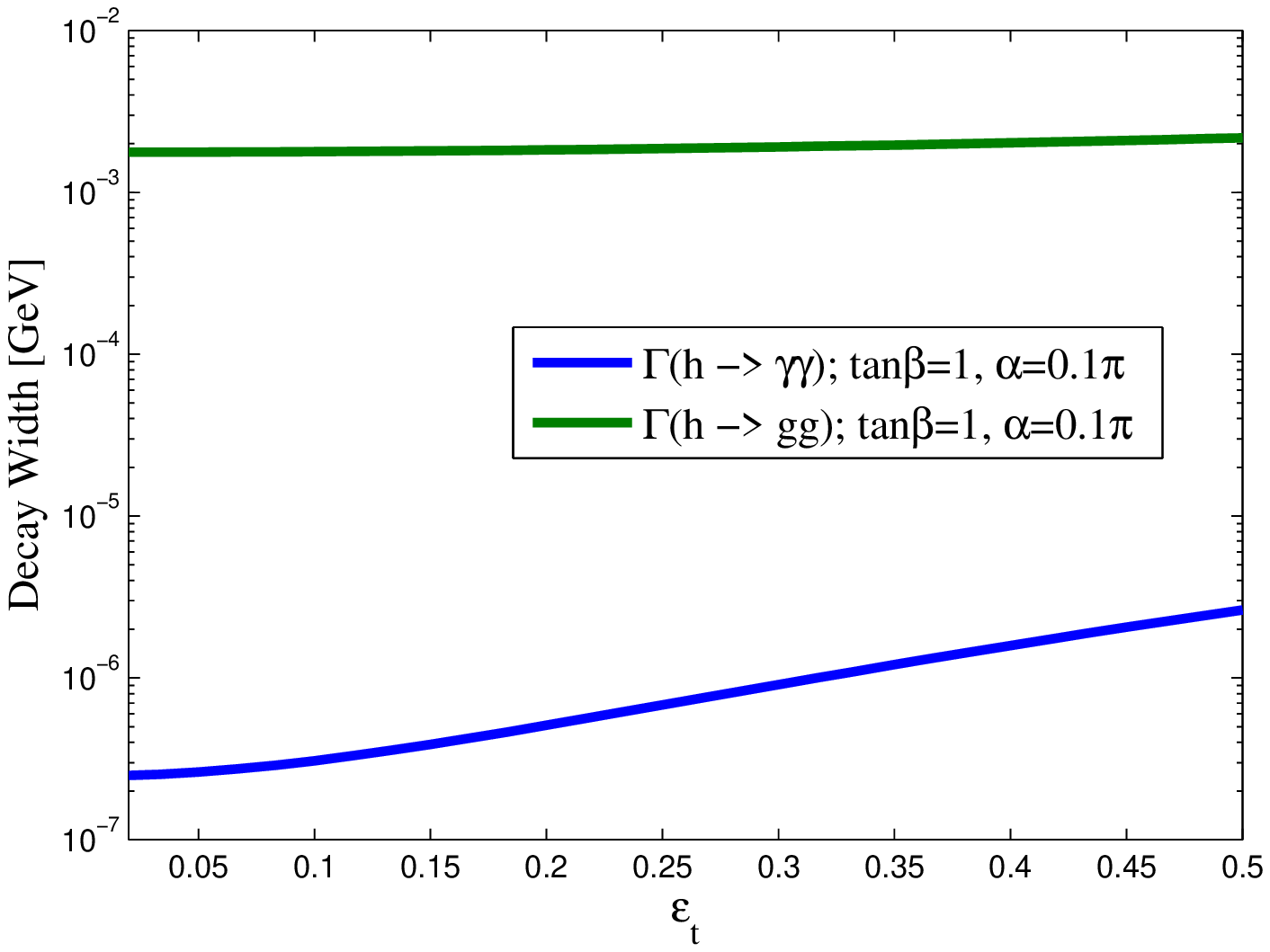,height=6.0cm,width=6.5cm,angle=0}
\vspace{-0.8cm}
\end{array}$
\end{center}
\caption{\emph{$\Gamma(h \to \gamma \gamma)$ and $\Gamma(h \to gg)$, as a function
of $\alpha$, $\tan{\beta}$ and $\epsilon_t$, for some representative values
of these parameters (as indicated in the plots)
and with $M_{4G} \equiv m_{t'}=m_{b'}=m_{l_4}=m_{\nu_4}=400$ GeV. \label{hgg}}}
\end{figure}

Clearly, once a new Higgs doublet is introduced, the phenomenology of the
Higgs particles production and decays becomes more complicated.
In particular, the new Yukawa couplings depend on several more parameters (i.e.,
in the 4G2HDM on $\epsilon_t$, $\epsilon_b$, $\tan{\beta}$ and $\alpha$)
and the CP-even Higgs couplings to the $W$ and to the $Z$ bosons
have extra pre-factors of $\sin \left(\alpha-\beta \right)$
and $\cos \left(\alpha-\beta \right)$ (the pseudoscalar $A$ does not
couple at tree-level to the $W$ and the $Z$).
As a result, the one-loop $h \to \gamma \gamma$ decay and the
leading $gg \to h$ Higgs production mechanism can be significantly altered compared to their
SM and SM4 values, depending on $\epsilon_t$, $\tan{\beta}$, $\alpha$
and on the 4th generation fermion masses (i.e.,
assuming $\epsilon_b << \epsilon_t$, therefore setting $\epsilon_b=0$
throughout our analysis).
This is demonstrated in Fig.~\ref{hgg}, where we plot
the widths $\Gamma(h \to \gamma \gamma)$ and $\Gamma(h \to gg)$, as a function
of these three parameters setting $M_{4G} \equiv m_{t'}=m_{b'}=m_{l_4}=m_{\nu_4}=400$ GeV.
The dependence on $\tan\beta$ is depicted in a narrow range around $\tan\beta \sim 1$,
for which the 4G2HDM is consistent with both EWPD \cite{4G2HDM}
and with the observed 125 GeV Higgs signals (see below).

\begin{figure}[htb]
\begin{center}
\epsfig{file=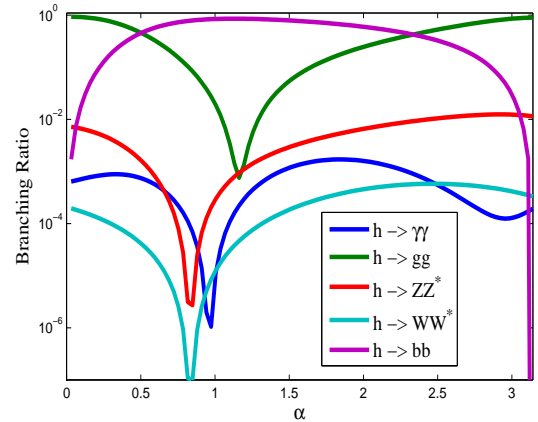,height=6cm,width=8cm,angle=0}
\vspace{-0.8cm}
\end{center}
\caption{\emph{The relevant branching ratios of $h$ in the 4G2HDM,
as a function of $\alpha$,
with $m_h=125$ GeV, $M_{4G}=400$ GeV, $\epsilon_t=0.5$ and $\tan\beta=1$.}
\label{Br_H}}
\end{figure}

We see that both $h \to \gamma \gamma$
and $h \to gg$ have a strong dependence on the Higgs mixing angle $\alpha$,
while $h \to \gamma \gamma$ is also very sensitive to
$\tan\beta$ and to the new $t - t^\prime$ mixing parameter
$\epsilon_t$, due to their role in the interference between the
fermion loops and the W-boson loop.
In Fig.~\ref{Br_H} we further plot the various relevant branching ratios
of a 125 GeV $h$ in the 4G2HDM, as a function of $\alpha$
for $\epsilon_t=0.5$, $\tan \beta=1$ and $M_{4G}=400$ GeV.

Let us now turn to the recently reported LHC Higgs searches and the implications
of the discovery of a 125 GeV Higgs-like particle on the 4G2HDM setup with a 4th generation
of fermions.
The quantity that is usually being used for comparison between the LHC and Tevatron
results and the expected signals in
various models is the normalized cross-section:
 \begin{eqnarray}
 R^{Model(Obs)}_{XX}=\frac{\sigma \left(pp/p\overline{p} \to h \to XX\right)_{Model(Obs)}}{\sigma \left(pp/p\overline{p} \to h \to XX\right)_{SM}} ~.
 \end{eqnarray}

For the observed ratios of cross-sections, i.e., the signal strengths $R^{Obs}_{XX}$,
and the corresponding errors $\sigma_{XX}$, we use the latest results as published in \cite{CMSHiggs,ATLASHiggs,TevatronHiggs}:
\begin{itemize}
 \item $VV \to h \to \gamma \gamma$: $2.2\pm 1.4$ (taken from $\gamma \gamma+2j)$
 \item $gg \to h \to \gamma \gamma$: $1.68\pm 0.42$
 \item $gg \to h \to W W^*$: $0.78\pm 0.3$
 \item $gg \to h \to Z Z^*$: $0.83\pm 0.3$
 \item $gg \to h \to \tau \tau$: $0.2\pm 0.85$
 \item $pp/p\overline{p} \to hW \to b\overline{b} W $: $1.8\pm 1.5$
\end{itemize}

\begin{figure*}[htb]
\begin{center}$
\begin{array}{cc}
\epsfig{file=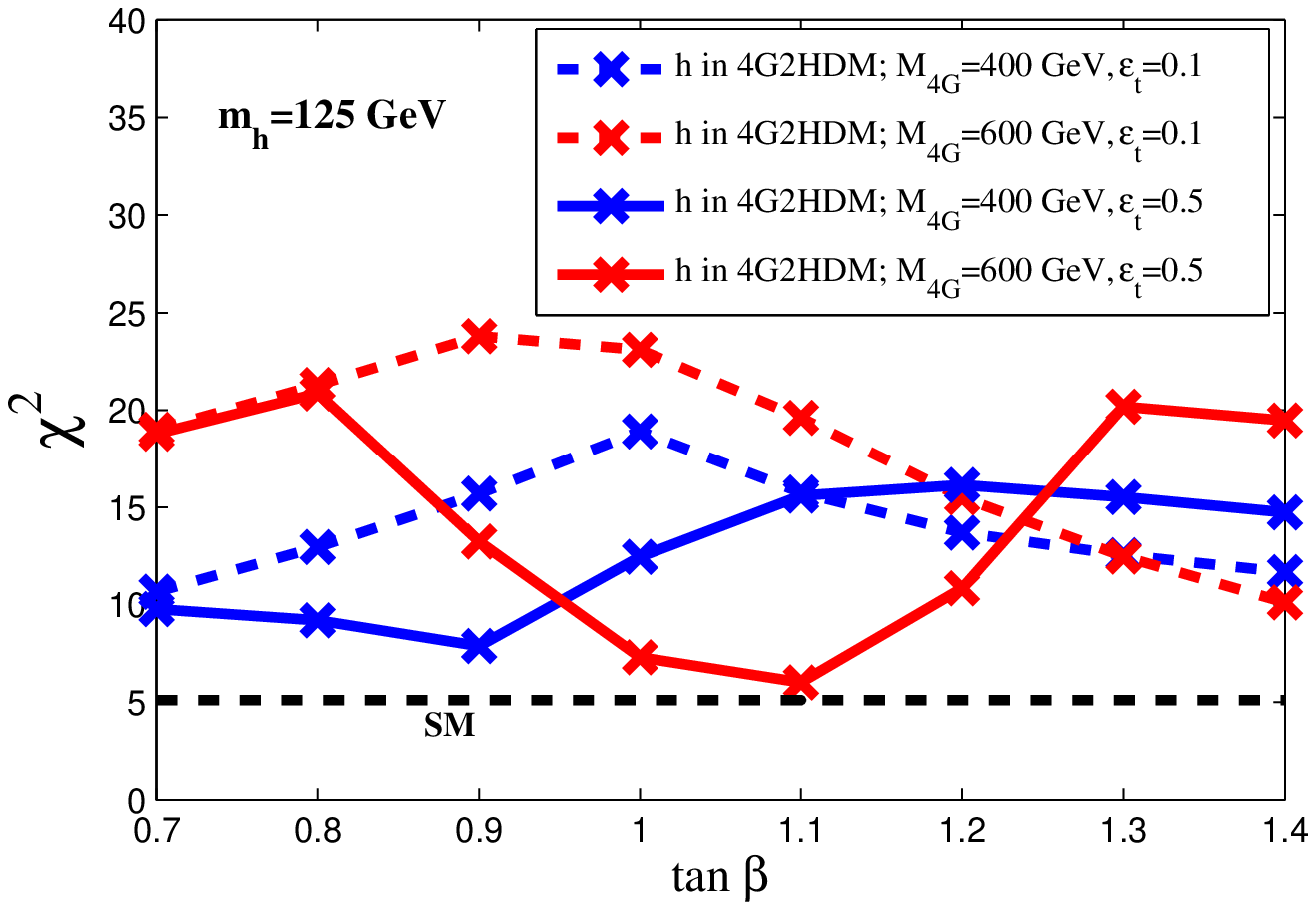,height=6cm,width=8.25cm,angle=0}
\epsfig{file=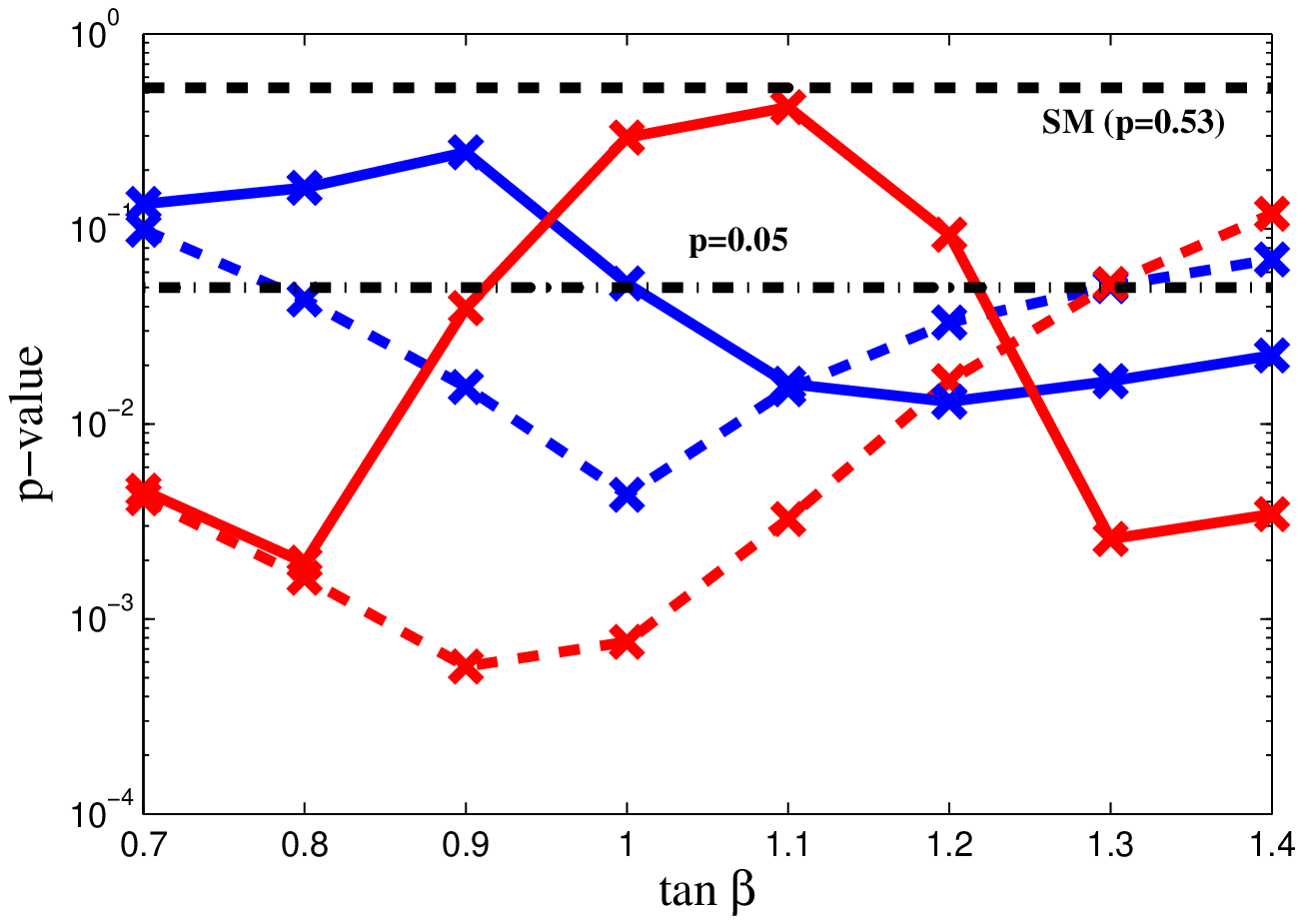,height=6cm,width=8.25cm,angle=0}
\vspace{-0.8cm}
\end{array}$
\end{center}
\caption{\emph{$\chi^2$ (left plot) and p-values (right plot), as a function of $\tan\beta$,
for the lightest 4G2HDM CP-even scalar $h$, with $m_h=125$ GeV, $\epsilon_t =0.1$ and $0.5$ and
$M_{4G} \equiv m_{t'}=m_{b'}=m_{l_4}=m_{\nu_4}=400$ and $600$ GeV.
The value of the Higgs mixing angle $\alpha$ is
the one which minimizes $\chi^2$ for each value of $\tan\beta$.
The SM best fit is shown by the horizontal dashed-line and the dash-doted line in the right plot
corresponds to $p=0.05$ and serves as a reference line. \label{ChiSqAll1}}}
\end{figure*}
\begin{figure*}[htb]
\begin{center}$
\begin{array}{cc}
\epsfig{file=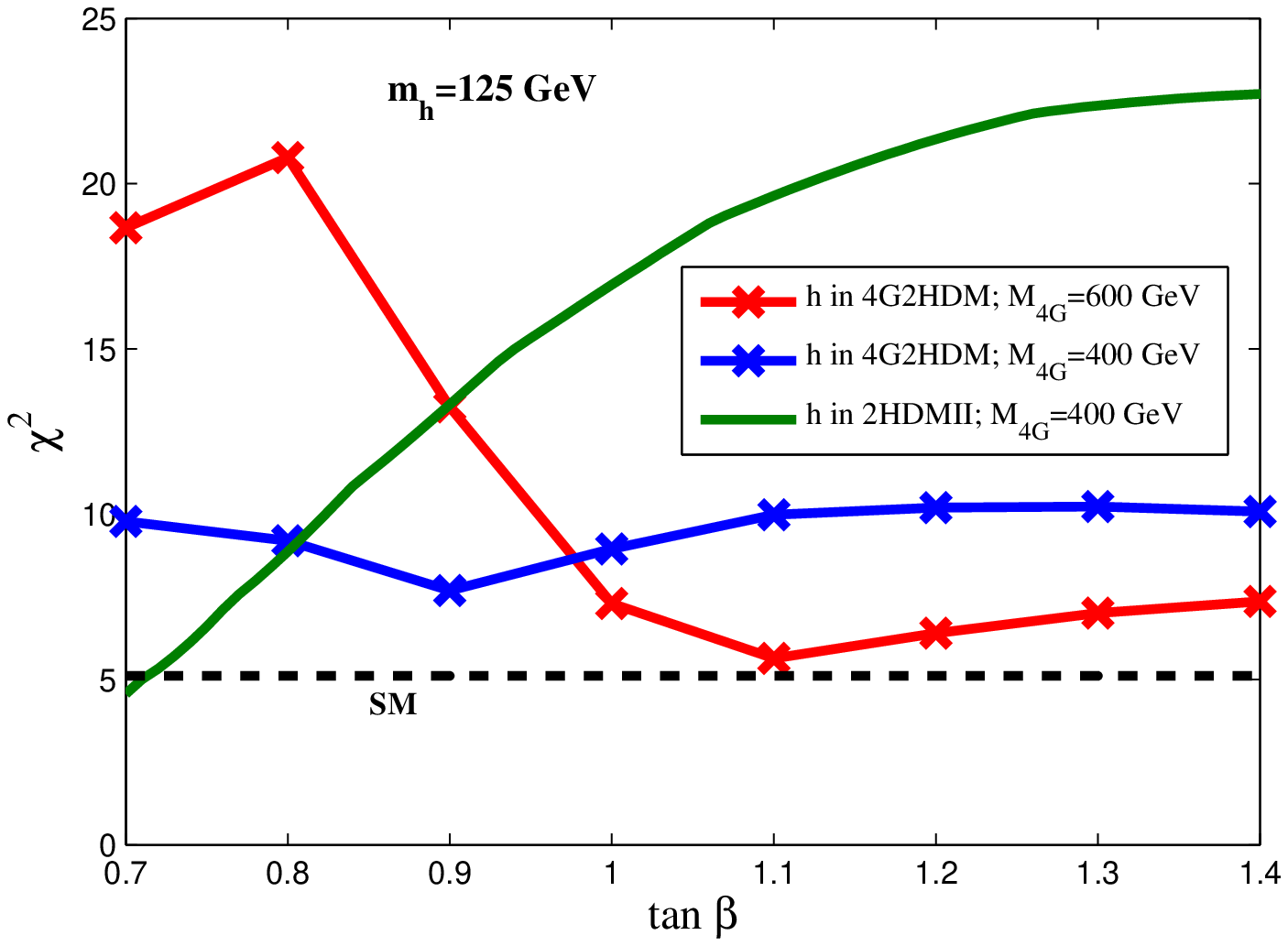,height=6cm,width=8.25cm,angle=0}
\epsfig{file=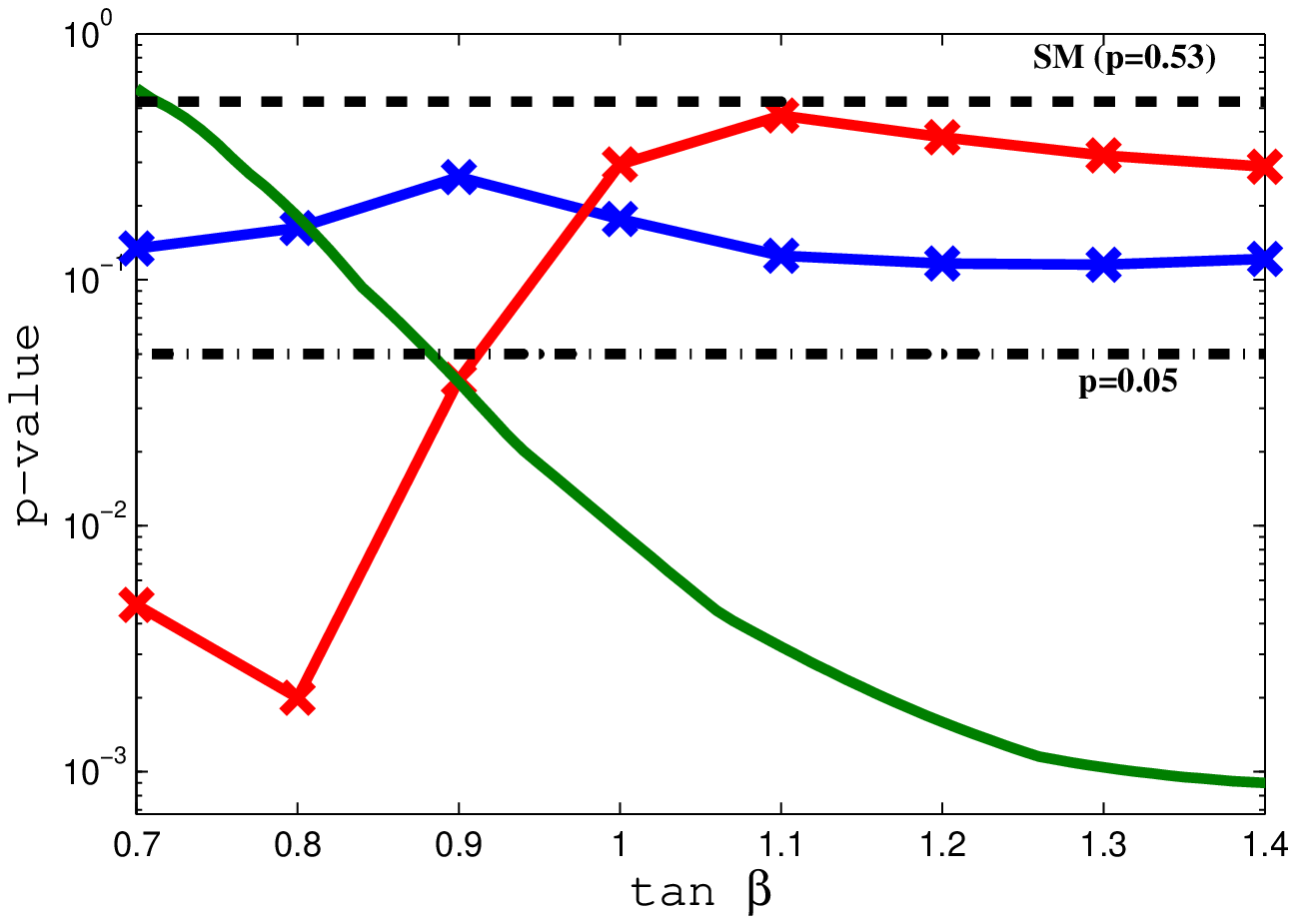,height=6cm,width=8.25cm,angle=0}
\vspace{-0.8cm}
\end{array}$
\end{center}
\caption{\emph{Same as Figure \ref{ChiSqAll1},
where here we minimize with respect to both $\epsilon_t$ and $\alpha$
for each value of $\tan\beta$.
Also shown are the $\chi^2$ and p-values
for a $125$ GeV Higgs in the SM and in
the type II 2HDM with a 4th generation of fermions (denoted by 2HDMII).}
\label{ChiSqAll2}}
\end{figure*}

The values given above are the result of a combination of the most recent data in each
channel.$^{[1]}$\footnotetext[1]{We combine the results
from the CMS and ATLAS experiments (for $pp/p\overline{p} \to hW \to b\overline{b} W $
we combine the results from CMS and Tevatron), where in cases where the measured value
was not explicitly given we estimate it from the published plots.}
The uncertainties are calculated by treating the reported experimental uncertainties as statistical and assigning 15\% theoretical uncertainty to the gluon fusion production mechanism and 5\% theoretical uncertainty on electroweak production mechanisms and on the branching fractions \cite{uncertainty}.
One can easily notice that the channels which have the highest sensitivity to the Higgs signals and contributed the most to the recent 125 GeV Higgs discovery are
$h\to \gamma \gamma$ and $h\to ZZ^*,WW^*$. In all other channels the results are not
conclusive and at this time they are consistent with the background only
hypothesis at the level of less than $2\sigma$.

Clearly, a SM Higgs is ideally most consistent with $R^{Obs}_{XX}=1$ in every channel, while in other models we expect some deviations in the various measured channels, depending on the parameters of the model and on the mass of the scalar candidate which should be compatible with the LHC results.
Thus, the comparison to any given model can be performed using a $\chi^2$ fit:
\begin{equation}
\chi^2=\sum_X{\frac{\left(R^{Model}_{XX}-R^{Obs}_{XX}\right)^2}{\sigma_{XX}^2}}~,
\end{equation}
 where $\sigma_{XX}$ are the errors on the observed cross-sections and
 $R^{Model}_{XX}$ is the calculated normalized cross-section in any given model.
 In particular, we take advantage of the fact that
 $\frac{\sigma\left(YY \to h \right)_{Model}}{\sigma\left(YY \to h \right)_{SM}}=\frac{\Gamma\left(h\to YY \right)_{Model}}{\Gamma\left(h\to YY \right)_{SM}}$, and calculate $R^{Model}_{XX}$ using
 \begin{eqnarray}
 R^{Model}_{XX}=\frac{\Gamma \left(h\to YY \right)_{Model}}{\Gamma \left(h\to YY \right)_{SM}} \cdot \frac{Br \left(h\to XX \right)_{Model}}{Br \left(h\to XX \right)_{SM}}~,
  \end{eqnarray}
 where $YY\to h$ is the Higgs production mechanism, i.e., either by gluon fusion $gg\to h$, vector boson fusion $WW/ZZ \to h$ or the associated
  Higgs-W production $W^*\to hW$ at the Tevatron.

The Higgs signals in a 2HDM setup with a 4th generation of fermions have already been discussed to some extent in the literature \cite{Bern,Gunion,valencia,He2,new2HDM4G}, but with no
general picture of how these signals match all the observed Higgs cross-sections reported above.
Here, we try to quantify how well the 2HDM scenarios (where the lightest Higgs particle, $h$,
has a mass of 125 GeV) fit all the available Higgs data,
by calculating the $\chi^2$ for all the relevant channels in two 2HDM
realizations with four generations - the 2HDMII and the 4G2HDM (see section \ref{sec2}).

We use the latest version of Hdecay \cite{Hdecay}, with
recent NLO contributions which also
include the heavy 4th generation fermions,
where we have inserted all the relevant couplings of the 4G2HDM
and the 2HDMII frameworks described in section \ref{sec2}.
For the 4th generation fermion masses involved in the loops of the decays $h \to VV$
(i.e., in the 1-loop NLO corrections for the cases $h \to ZZ^*,WW^*$),
we have used the approximation of
a degenerate 4th generation fermion sector, where we have tested below two representative cases: $m_{t^\prime}=m_{b^\prime}=m_{l_4}=m_{\nu_4} \equiv M_{4G}= 400$ and $600$ GeV (the effect
of mass splittings between 4th generation fermions on $\Gamma(h \to VV)$ is negligible).
It is important to note that, while the first case ($M_{4G}=400$ GeV) is excluded for the SM4 \cite{LHC_4G}, it is not necessarily excluded for the 4G2HDM, since in this model the decay patterns of $t^\prime$ and $b^\prime$ can have a completely different topology, e.g., having $BR\left(t'\to t h \right) \sim 1$, for which the current limits
(which are based on the ``standard" SM4 decays $t^\prime \to bW$ and $b^\prime \to tW$)
do not apply, see \cite{Geller}.
\begin{figure}[htb]
\begin{center}
\epsfig{file=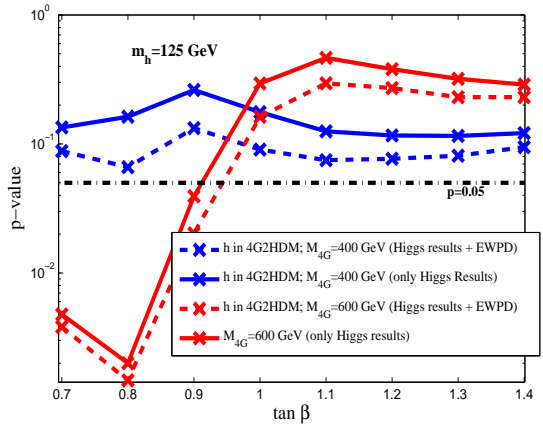,height=6cm,width=8cm,angle=0}
\vspace{-0.8cm}
\end{center}
\caption{\emph{A comparison between the p-values for the compatibility of
the $125$ GeV $h$ of the 4G2HDM
with the Higgs search results, with (dashed lines) and without (solid lines)
imposing the constraints from
EWPD as given in \cite{4G2HDM,AHEP-rev}. The dash-doted line corresponds to $p=0.05$.
Here also, the parameters $\alpha$ and $\epsilon_t$ are chosen by a minimization of the
$\chi^2$.} \label{HiggsEwpd}}
\end{figure}

As mentioned earlier, we find that the simple SM4 case, with a 125 GeV Higgs is
excluded to many $\sigma$'s when confronted with the Higgs search results.
Also, as was already noted in \cite{Gunion} in the context of
the ``standard" 2HDMII (i.e., with four generations),
we find that the simplest case of a light 125 GeV pseudoscalar $A$ of any 2HDM,
with and without a 4th family, is not compatible with the Higgs data,
irrespective of the 4th generation fermion masses.
In particular, the
signals of the 125 GeV Higgs decaying into a pair of vector bosons, $h \to ZZ$ and
$h \to WW$,
excludes this possibility due to the absence of tree-level
$AZZ$ and $AWW$ couplings. We, therefore, focus below only on the case
where the observed 125 GeV Higgs-like particle is the
lighter CP-even Higgs, $h$.

We plot in Fig.~\ref{ChiSqAll1}
the resulting $\chi^2$ and p-values
in the 4G2HDM case
(combining all the six reported Higgs decay channels above),
with $m_h=125$ GeV, $M_{4G}=400$ and 600 GeV, $\epsilon_t = 0.1$ and 0.5 and
for $0.7<\tan \beta<1.4$ (this range is allowed by EWPD and flavor physics in the
2HDM 4th generation setups, see \cite{4G2HDM,AHEP-rev}).
The value of the Higgs mixing angle $\alpha$ is the one
which minimizes the $\chi^2$ for each value of $\tan\beta$.
The SM best fit is also shown in the plot.
In Fig.~\ref{ChiSqAll2} we further plot the resulting
$\chi^2$ and p-values as a function of $\tan\beta$,
this time minimizing for each
value of $\tan\beta$ with respect to both $\alpha$ and $\epsilon_t$ (in the 4G2HDM case).
For comparison, we also show in Fig.~\ref{ChiSqAll2} the $\chi^2$ and p-values for a
125 GeV $h$ in the 2HDMII with a 4th generation
and in the SM.

Looking at the p-values in Figs.~\ref{ChiSqAll1} and \ref{ChiSqAll2}
(which ``measure" the extent to which
a given model can be successfully used to interpret the Higgs data
in all the measured decay channels)
we see that, $h$ of the 4G2HDM with $\tan\beta \sim {\cal O}(1)$ and
$M_{4G}=400-600$ GeV is a
good candidate for the recently observed 125 GeV Higgs,
giving a fit comparable to the SM fit.
The ``standard" 2HDMII setup with $M_{4G} = 400$ GeV
is also found to be consistent with the Higgs data
in a narrower range of $\tan\beta \lsim 0.9$.
We find that the fit favors a large $t - t^\prime$ mixing
parameter $\epsilon_t$, implying ${\rm BR}(t^\prime \to th) \sim {\cal O}(1)$
which completely changes the $t^\prime$ decay pattern
\cite{4G2HDM} and, therefore, significantly
relaxing the current bounds on $m_{t^\prime}$ \cite{Geller}.
This can be seen
in Table \ref{tab1} where we list 6 representative sets of best fitted values
(to be used in the plots below) for $\left\{ \tan\beta, \alpha, \epsilon_t , M_{4G} \right\}$
in the 4G2HDM, that correspond to points on the best fitted 4G2HDM curves shown in Fig.~\ref{ChiSqAll2}.

In Fig.~\ref{HiggsEwpd} we further test the goodness of fit for the 125 GeV $h$ of the 4G2HDM,
where, in addition to the Higgs results, we explicitly imposed
the constraints on the 4G2HDM parameter space from EWPD (from the S and T parameters
and from $Z \to b \bar b$) using the results in \cite{4G2HDM,AHEP-rev}. Evidently,
our conclusions above do not change after adding the EWPD constraints to the analysis.

Finally, we note that we have also tested the 3 generations
type II 2HDM and found that its lightest CP-even Higgs
is also a good candidate for the observed
125 GeV Higgs particle, giving a fit which is also
comparable to the SM fit for $\tan\beta \sim {\cal O}(1)$.

\begin{table}[htb]
\begin{center}
\begin{tabular}{c|c|c|c|c}
Point \# & $\tan\beta$ & $\alpha$ & $\epsilon_t$ & $M_{4G}$ [GeV] \\
\hline \hline
P1 & 0.7 & 0.09$\pi$ & 0.5 & 400 \\
P2 & 0.7 & 0.51$\pi$ & 0.433 & 600 \\
P3 & 1.0 & 0.1$\pi$ & 0.42 & 400 \\
P4 & 1.0 & 0.08$\pi$ & 0.5 & 600 \\
P5 & 1.3 & 0.11$\pi$ & 0.3 & 400 \\
P6 & 1.3 & 0.07$\pi$ & 0.33 & 600
\end{tabular}
\caption{\emph{Six representative best fitted sets of values for
$\left\{ \tan\beta, \alpha, \epsilon_t , M_{4G} \right\}$
in the 4G2HDM, corresponding to  points on the best fitted 4G2HDM curves shown in Fig.~\ref{ChiSqAll2}.}}
\label{tab1}
\end{center}
\end{table}

\section{Higgs phenomenology in the 4G2HDM}

\begin{figure}[ht]
\begin{center}
$
\begin{array}{cc}
\epsfig{file=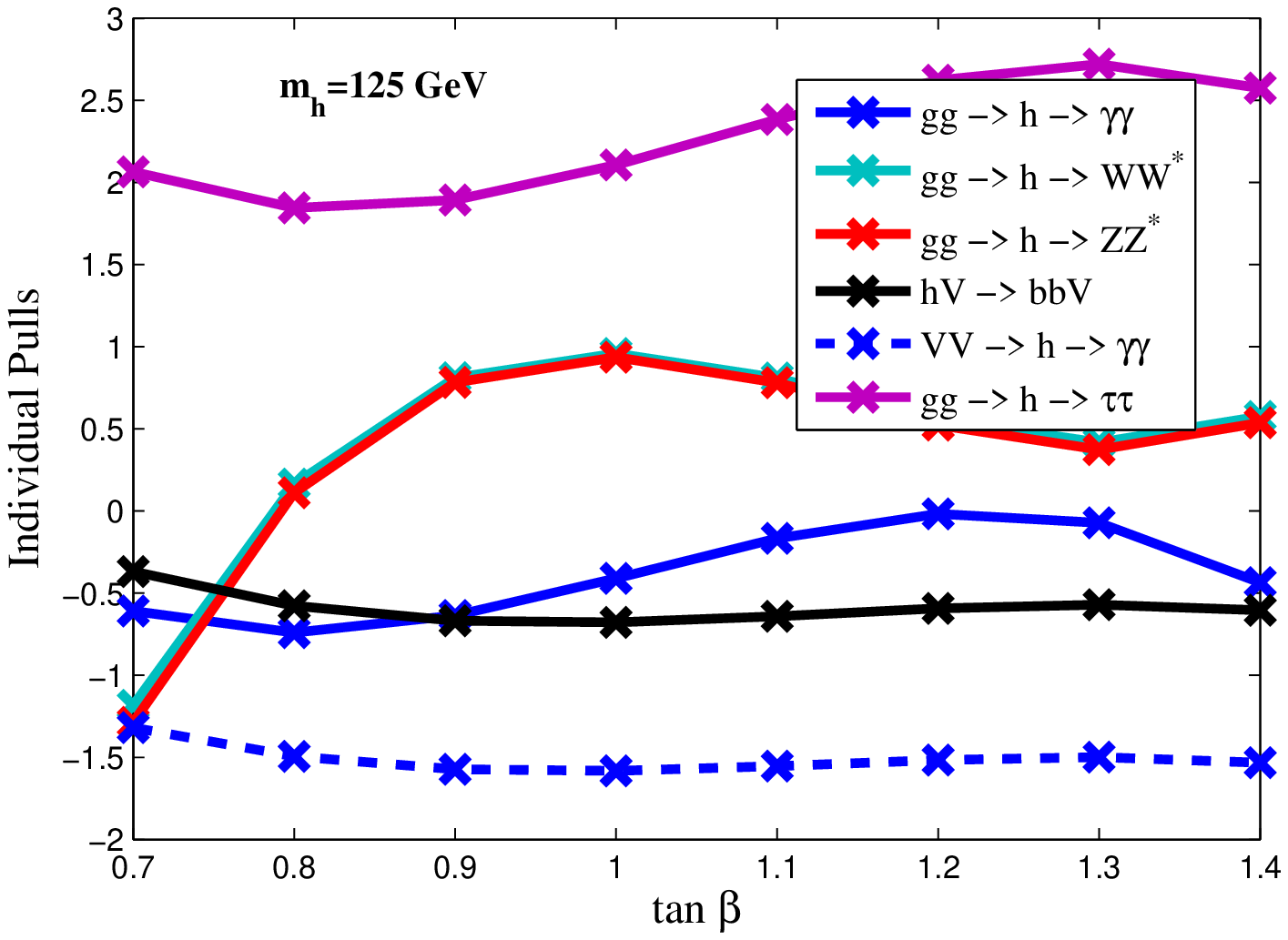,height=6cm,width=8.25cm,angle=0}\\
\epsfig{file=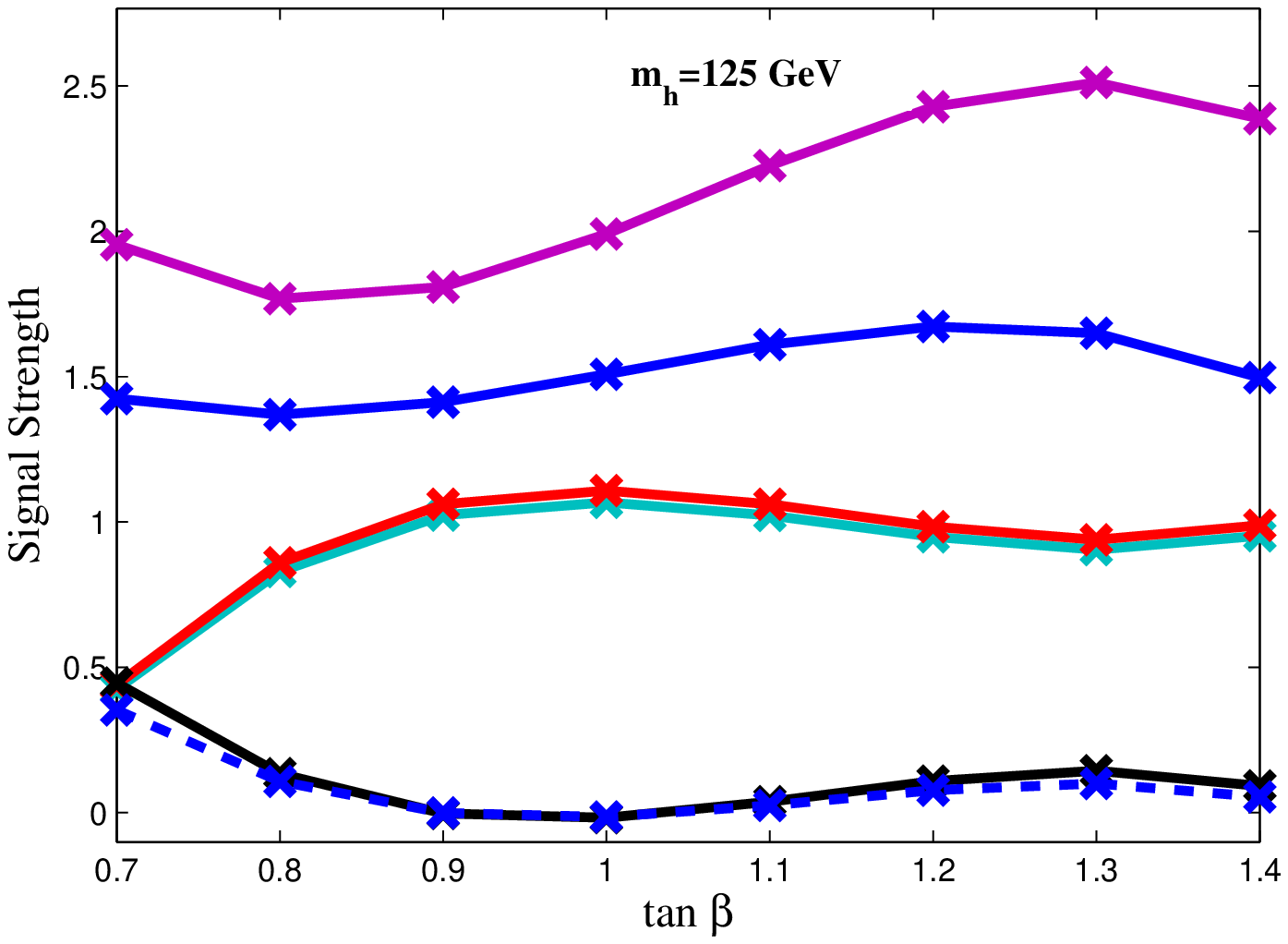,height=6cm,width=8.25cm,angle=0}
\vspace{-0.8cm}
\end{array}$
\end{center}
\caption{\emph{The individual pulls
$\frac{\left(R^{Model}_{XX}-R^{Obs}_{XX}\right)}{\sigma_{XX}}$
(upper plot), and the
signal strengths $R^{Model}_{XX}$ (lower plot), in the different channels,
that correspond to the best fitted 4G2HDM curve
with $m_h=125$ GeV and $M_{4G}=400$ GeV,
shown in Fig.~\ref{ChiSqAll2}.}
\label{indv_pulls}}.
\end{figure}

In Fig.~\ref{indv_pulls} we plot the individual pulls
and the signal strengths for the various measured channels,
$\left(R^{4G2HDM}_{XX}-R^{Obs}_{XX}\right)/\sigma_{XX}$ and
$R^{4G2HDM}_{XX}$, respectively,
as a function of $\tan\beta$,
for the above best fitted 4G2HDM curve with
$M_{4G}=400$ GeV.
We see that appreciable deviations from the SM are expected in the channels
$gg \to h \to \tau \tau$, $VV \to h \to \gamma \gamma$ and $h V \to bb V$.
In particular,
the most notable effects
are about a $1.5 \sigma$ deviation (from the observed value) in the VBF diphoton channel
$VV\to h \to \gamma\gamma$ and a $2-2.5 \sigma$ deviation in the
$gg \to h \to \tau \tau$ channel. The deviations in these channels
are in fact a prediction of the 4G2HDM strictly based on the current Higgs data,
which could play a crucial role as data
with higher statistics becomes available. They can be understood as follows:
the channels that dominate the fit (i.e., having a higher statistical
significance due to their smaller errors) are $gg \to h \to \gamma \gamma,ZZ^*,WW^*$.
Thus, since the $gg \to h$ production vertex
is generically enhanced by the $t^\prime,b^\prime$ loops,
the fit then searches for values of the relevant 4G2HDM parameters which
decrease the $h \to \gamma \gamma,ZZ^*,WW^*$ decays in the appropriate
amount. This in turn leads to an enhanced $gg \to h \to \tau \tau$
(i.e., due to the enhancement in the $gg \to h$ production vertex)
and to a decrease in the $VV \to h \to \gamma \gamma$ and
$p \bar p/pp \to W \to h W \to bb W$, which are
independent of the enhanced $ggh$ vertex
but are sensitive to the decreased $VVh$ one.
It is important to note that some of the characteristics of these 
``predictions" can change with more data collected. 
\begin{figure}[htb]
\begin{center}
\epsfig{file=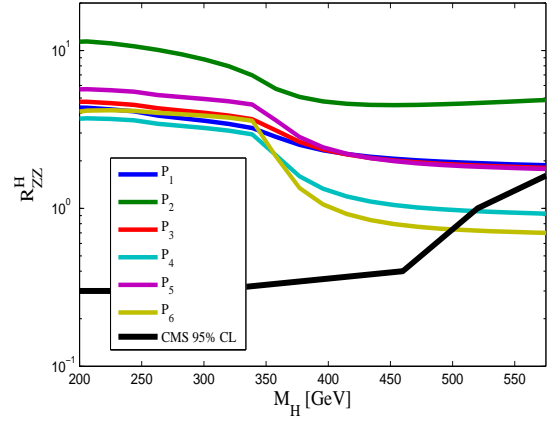,height=6cm,width=8cm,angle=0}
\vspace{-0.4cm}
\end{center}
\caption{\emph{The signal strength in the $H\to ZZ$ channel, for
the 6 best fitted sets of values in Table \ref{tab1}.
Also shown is the approximate observed CMS limit on the signal strength in
the $ZZ$ channel, i.e., $R_{ZZ}^{Obs}$, see also text.} \label{ScalarH}}
\end{figure}
\begin{figure}[htb]
\begin{center}
\epsfig{file=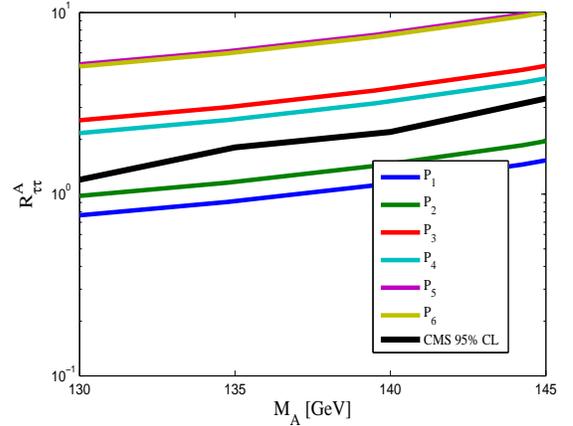,height=6cm,width=8cm,angle=0}
\vspace{-0.4cm}
\end{center}
\caption{\emph{The signal strengths in the $A\to \tau \tau$ channel,
for
the 6 best fitted sets of values in Table \ref{tab1}.
Also shown is the approximate observed CMS limit on signal strengths in the
$\tau \tau$ channel, i.e., $R_{\tau \tau}^{Obs}$, see also text.}
\label{ScalarAtau}}
\end{figure}
\begin{figure}[htb]
\begin{center}
\epsfig{file=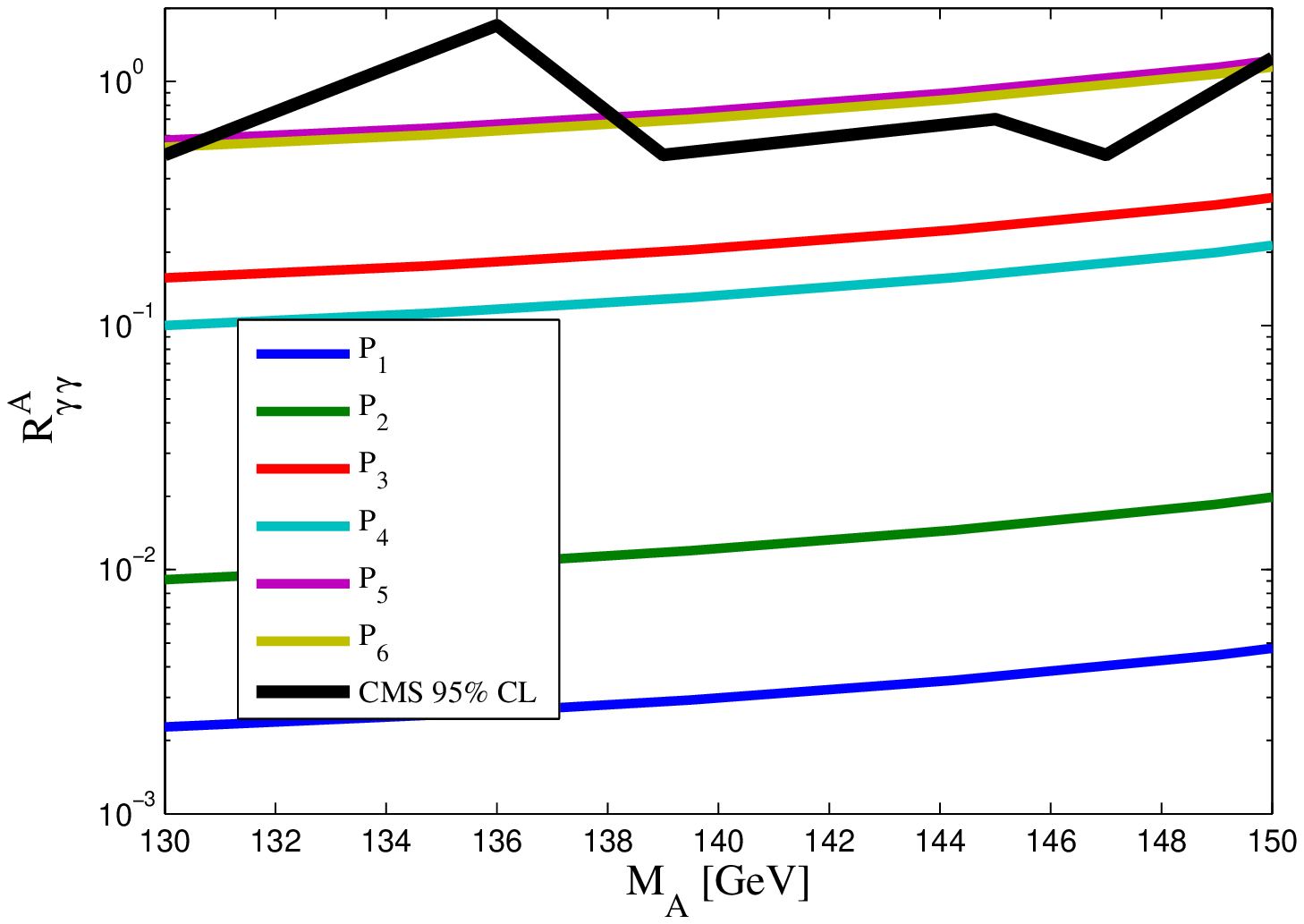,height=6cm,width=8cm,angle=0}
\vspace{-0.4cm}
\end{center}
\caption{\emph{The signal strengths in the $A\to \gamma \gamma$ channel,
for
the 6 best fitted sets of values in Table \ref{tab1}.
Also shown is the approximate observed CMS limit on signal strengths in the
$\gamma \gamma$ channel, i.e., $R_{\gamma \gamma}^{Obs}$, see also text.}
\label{ScalarAgam}}
\end{figure}

We conclude with the implications of the above results for the
other two neutral scalars of the 4G2HDM.
For the heavier CP-even neutral Higgs, $H$,
we consider its decays to $ZZ$ and $WW$, which are currently
the most sensitive channels in which searches for a heavy SM Higgs
were performed at the LHC. A useful approximation of the expected
exclusion range on $m_H$ can be performed by comparing
the calculated signal strengths:
\begin{equation}
R_{ZZ/WW}^{H} \equiv \frac{\sigma(pp \to H \to ZZ/WW)_{4G2HDM}}{\sigma(pp \to H \to ZZ/WW)_{SM}}
~,
\end{equation}
to the observed/measured values of this quantity, i.e., to $R_{ZZ/WW}^{Obs}$
(note that in the 4G2HDM we find $R_{WW}^{H} \sim R_{ZZ}^{H}$ for $m_H > 2m_W$).
In Fig.~\ref{ScalarH} we plot $R_{ZZ}^{H}$ as a function of $m_H$ for the 6
best fitted points
of the relevant 4G2HDM parameter space, given in Table \ref{tab1}.
We also show in Fig.~\ref{ScalarH} an approximate exclusion line for
$R_{ZZ}$, i.e., for the observed signal strength $R_{ZZ}^{Obs}$, which we have extracted
from the
most recent CMS exclusion plot in this channel (see \cite{CMSZZ}) and
which is currently the most stringent observed exclusion limit for
a heavy Higgs with a mass $\gsim 200$ GeV.
We see that
$m_H \lsim 600$ GeV is excluded
by the current data in the $H \to ZZ$ channel
for points P1, P2, P3 and P5 (i.e., $R_{ZZ}^{H}(P1,P2,P3,P5) > R_{ZZ}^{Obs}$ for
$m_H \lsim 600$ GeV),
while for point P4 and P6 $m_H \gsim 500$ GeV is allowed.

The current CMS and ATLAS Higgs data in the $ZZ$ and $WW$ channels are not sensitive
to the pseudoscalar $A$, due to the absence of a tree level
$AZZ$ and $AWW$ coupling (and due to
the smallness of the corresponding $AZZ$ and $AWW$ one-loop couplings \cite{gunionAZZ}).
Therefore, the only relevant search channels which are currently sensitive to
the $A$ decays are $A \to \gamma \gamma$ and $A \to \tau \tau$, for
which a search for the Higgs was performed up to a Higgs mass slightly below $2 m_W$
by both CMS and ATLAS.
Defining the signal strengths for the $A$ signals as:
\begin{equation}
R_{\tau \tau/\gamma \gamma}^{A} \equiv \frac{\sigma(pp \to A \to \tau \tau/\gamma \gamma)_{4G2HDM}}{\sigma(pp \to H \to \tau \tau / \gamma \gamma)_{SM}}
~,
\end{equation}
we plot in Figs.~\ref{ScalarAtau} and \ref{ScalarAgam}
$R_{\tau \tau}^{A}$ and $R_{\gamma \gamma}^{A}$, respectively,
as a function of $m_A$ (we assume that $m_A > m_h$),
for points P1-P6 of Table \ref{tab1}.
Here also, we
plot the existing approximate exclusion lines
$R_{\tau \tau}^{Obs}$ and $R_{\gamma \gamma}^{Obs}$, based
on the most recent CMS analysis,
which currently gives the most stringent limits in these
two channels \cite{CMStt,CMSgg}.
We see that a pseudoscalar as light as 130 GeV is allowed
by the current data, e.g., for for points
P1 and P2.

\section{Summary}

We have studied the recently measured Higgs signals in the framework of
a specific 2HDM with a fourth generation of fermions (the 4G2HDM suggested in \cite{4G2HDM}),
designed and motivated
by the possibility that the sub-TeV Higgs particles are condensates of the heavy 4th generation
fermions, which are therefore, viewed as the agents of dynamical electroweak symmetry breaking.

We find that the lightest CP-even Higgs state of this model, $h$,
is a good candidate
for the recently discovered 125 GeV Higgs signals in all the measured
channels, within a large portion of the 4G2HDM allowed parameter
space, which is consistent with the current bounds from EWPD.
In particular, for typical 4th generation fermion masses
in the range $M_{4G}=400-600$ GeV,
$\tan\beta \sim {\cal O}(1)$ and a large
$t - t^\prime$ mixing (the parameter $\epsilon_t$ predicted by the model),
the lightest 4G2HDM Higgs gives a good overall fit
to the current 125 GeV Higgs data - roughly comparable to the SM fit.

For these values of the 4G2HDM parameter space, in particular with $\epsilon_t \sim 0.5$,
the flavor changing $t^\prime$ decay $t^\prime \to t h$ dominates with
${\rm BR}(t^\prime \to t h) \sim 1$, leading to different  $pp \to t^\prime \bar t^\prime$
signatures (than the simple SM4) that can be searched for at the LHC using the methods suggested in \cite{Geller}.

We also find that, {\it based on the current Higgs data}, the 4G2HDM
predicts large deviations from the SM in the channels
$pp \to h \to \tau \tau$, $VV \to h \to \gamma \gamma$ and
$hV \to V bb$, which remains to be tested with more data.

Finally,
the heavier CP-even Higgs state, $H$, is found to be excluded in this model up to $\sim 500$ GeV,
while the pseudoscalar Higgs state, $A$, can be as light as 130 GeV and can, therefore, be discovered
(or ruled out in this small mass range) with more data collected in the $pp \to \tau \tau,\gamma \gamma$ channels.

\bigskip

{\bf Acknowledgments:} SBS and MG acknowledge research support from the Technion.
GE thanks R. Godbole and S. Vempati for hospitality and discussions.
The work of AS was supported in part by the U.S. DOE contract
\#DE-AC02-98CH10886(BNL).

\end{document}